# Has the impact flux of small and large asteroids varied through time on Mars, the Earth and the Moon?

Anthony Lagain[a,*], Mikhail Kreslavsky[b], David Baratoux[c,d], Yebo Liu[e], Hadrien Devillepoix[a], Philip Bland[a], Gretchen K. Benedix[a,f,g], Luc S. Doucet[e], Konstantinos Servis[h],

[a]*Space Science and Technology Centre, School of Earth and Planetary Sciences, Curtin University, Kent St, Bentley, 6102, WA, Australia*
[b]*Earth and Planetary Sciences, University of California - Santa Cruz, CA, USA*
[c]*Géosciences Environnement Toulouse, University of Toulouse, 14, Avenue Edouard Belin, Toulouse, 31400, France*
[d]*University Félix Houphouët-Boigny, UFR des Scinces de la Terre et des Ressoures Minières, Cocody, Abidjan, Côte d'Ivoire.*
[e]*Earth Dynamics Research Group, The Institute for Geoscience Research (TIGeR), Department of Earth and Planetary Sciences, Curtin University, Kent St, Bentley, 6102, WA, Australia*
[f]*Planetary Sciences Institute, Tucson, AZ, USA*
[g]*Department of Earth and Planetary Sciences, Western Australian Museum, WA, Australia*
[h]*CSIRO - Pawsey Supercomputing Centre, WA, Australia*

[*]Corresponding author.
*Email address:* anthony.lagain@curtin.edu.au (Anthony Lagain)

## Abstract

The impact flux over the last 3 Ga in the inner Solar System is commonly assumed to be constant through time due to insufficient data to warrant a different choice for this range of time. However, asteroid break-up events in the main belt may have been responsible for cratering spikes over the last ~2 Ga on the Earth-Moon system. Due to its proximity with the main asteroid belt, i.e., the main impactors reservoir, Mars is at the outpost of these events with respect to the other inner planets. We investigate here, from automatic crater identification, the possible variations of the size frequency distributions of impactors from the record of small craters of 521 impact craters larger than 20 km in diameter. We show that 49 craters (out of the 521) correspond to the complete crater population of this size formed over the last 600 Ma. Our results on Mars show that the flux of both small (> 5 m) and large asteroids (> 1 km) are coupled, does not vary between each other over the last 600 Ma. Existing data sets for large craters on the Earth and the Moon are analyzed and compared to our results on Mars. On Earth, we infer the formation location of a set of impact craters thanks to plate tectonic reconstruction and show that a cluster of craters formed during the Ordovician period, about 470 Ma ago, appears to be a preservation bias. On the Moon, the late increase seen in the crater age signal can be due to the uncertain calibration method used to date those impacts (i.e. rock abundance in lunar impact ejecta), and other calibrations are consistent with a constant crater production rate. We conclude to a coupling of the crater production rate between kilometer-size craters (~100 m asteroids) and down to ~100 m in diameter (~5 m asteroids) in the inner Solar System. This is consistent with the traditional model for delivering asteroids to planet- crossing obits: the Yarkovsky effect slowly pushes the large debris from asteroid break-ups towards orbital resonances while smaller debris are grinded through collisional cascades. This suggests that the long-term impact flux of asteroids > 5 m is most likely constant over the last 600 Ma, and that the influence of past asteroid break-ups in the cratering rate for D > 100 m is limited or inexistent.

*Highlights*:

- Semi-automatic dating of 521 impact craters on Mars
- Statistical assessment of the lunar and terrestrial cratering rate
- Plate tectonic reconstruction of the formation location of terrestrial craters
- Coupling between the impact flux of small and large impactors
- The impact flux is constant over the last 600 Ma in the inner Solar System

---

:

1. Introduction

Mapping and counting impact craters is the most widely used tool to derive qualitative and quantitative temporal information on geological events and processes shaping the surface of terrestrial planets (e.g. Fassett (2016) and references therein). Radiometric ages of lunar samples from human and robotic missions have been combined with the crater density of the terrains where they were collected to calibrating absolute cratering chronologies (Hartmann and Neukum, 2001; Ivanov, 2001). These chronometers have then been adapted for Mars, Mercury and other solid bodies, including minor planets and icy satellites. Due to the lack of lunar samples covering recent epochs, i.e., the last 3 Ga, one of the most important sources of uncertainty of this dating method is the recent (0 – 3 Ga) evolution of the impact cratering flux. It is generally assumed a constant flux and a steady size-distribution of impactors. In other words, the recent production rate of impact craters on the surface of terrestrial planets and their Moon is considered to be the result of a set of physical processes for delivering impactors in the inner Solar System, that together form a stationary stochastic process.

As a consequence, impact crater age derivation on planetary surfaces (on the Moon and Mars) may be performed by crater counts. Nevertheless, recent studies argued for the existence of periods of a quiescent impact cratering ("lull"), alternating with increases in large crater production ("spike"), assuming the production of small craters remained constant. More specifically, using counts of small craters superposed on the floors of a set of 50 km and larger craters on the Moon, Kirchoff et al. (2021) reported a relative increase in flux occurring 2 Ga ago followed by a decrease 1 Ga ago. Using crater density on proximal ejecta of lunar craters greater than 20 km, Terada et al. (2020) argued for an impactor flux higher by a factor of 2-3 between

800 and 830 Ma.

Using the small crater populations to infer the formation rate of the large ones requires to make the assumption that the rate of small craters is constant to be able to use chronological models (Hartmann and Neukum, 2001; Hartmann, 2005). Actually, the discrepancy between the cratering rate of large craters inferred from the constant flux of small crater formation indicates nothing else that a decoupling between the small and large crater production rate, independently on the assumption made on the rate of small crater formation. In turn, it should be noted that no information about the absolute calibration of impact rate as a function of time may be obtained via these approaches. Absolute calibration of the impact rate needs independent data, such as absolute ages of geological events and samples by isotopic methods. With this in mind, Lagain et al. (2020) concluded that a non-steady-state size distributions of impactors in the main asteroid belt was responsible for the small crater counts observed on large impact craters on Mars over the last 2 Ga (i.e. a decoupling of the formation rate of craters smaller than 1 km and those larger than 1 km).

The fact that that no information about how long and when a potential fluctuation of the impact flux occurred can be derived with this technique can be illustrated through a simple case: let's consider a surface A exposed to the meteoritic bombardment for a period t at a constant impact rate with a steady impactor size-frequency distribution. At the end of the period, there is an observable crater population. Let's now consider a surface B, cratered for half of the period t with a rate double that of the surface A and then with a null cratering rate for the whole remaining time (i.e. t/2). At the end, we expect that the two surfaces A and B present a population statistically identical. Despite two different rate functions affecting the same surface, the crater populations observed at the end of the experience is the same because all that can be observed is the present time-integrated accumulation of a crater population. The same logic applies when the impact crater sizes vary. The impact rate for a given size is given by a time-dependent function multiplied by a production function (a size-frequency distribution). It is possible to use the small crater population to infer the ages of the large craters, and then obtain a rate of the large crater formation. What if a discrepancy is observed, i.e. the inferred rate for the formation of large craters from small crater counts differ from the expected one? This indicates that the hypothesis of a constant production function (or impactor size-frequency distribution) is not valid (see discussions in Lagain et al. (2020) and Mazrouei et al. (2019) supplementary materials). However, it is not possible to infer any information on the impact flux variation such the duration or the intensity of a spike or a lull period and when it occurred without other independent data. Moreover, any observed mismatch needs to be statistically assessed in order to distinguish variations compatible with a stationary stochastic process from non-stationary variations induced by an underlying process.

With this logic kept in mind, the present study aims first at investigating possible variations of the production function using a set of large craters on Mars with their record of small craters. Then, we will consider that the proximity of Mars from the main belt, as well as the Earth-Moon distance exclude the possibility that one of these three bodies experienced a cratering spike (or a lull) in their geological history whether the others did not. We will therefore discuss our martian data in lights of cratering record for the Moon and the Earth (independent data), to discuss potential fluctuation of the impact cratering rate in the inner solar system.

## 2. Data and methods

While large crater radiometric ages are well documented on Earth (e.g. Jourdan et al., 2009; Schmieder and Kring, 2020) and lunar crater ages from the Diviner rock abundance have been previously reported (Mazrouei et al., 2019), sets of large craters age on Mars exist but are limited in sample size, crater size and age range (Robbins et al., 2013; Lagain et al., 2020). In this section, we first present the terrestrial and lunar crater data we used to analyze the impact cratering rate in the Earth-Moon system, before presenting the method we developed to measure large impact craters model age on Mars.

*2.1. The terrestrial impact crater population*

The impact crater population on Earth is limited to 200 confirmed structures and is far from being complete over the Earth history (Hergarten and Kenkmann, 2015; Johnson and Bowling, 2014). Several factors bias the age frequency distribution of terrestrial impact craters including (1) the erosion and plate tectonic, (2) the heterogeneous efforts for geological exploration and mapping and the number of researchers involved in impact science across the world, (3) the ability to precisely date an impact event, mostly dependent to the type of the target (sedimentary, volcanic…) and the availability of suitable samples for dating, such as impact melt rocks or well-characterized sedimentary series (e.g. Bland, 2005).

These biases being difficult (and even impossible) to deconvolve, any attempt to derive the absolute impact flux evolution of large asteroids from the terrestrial impact cratering record seems to be doomed to failure.

Nevertheless, the biases mentioned above are less critical for the most recent periods of time.

Previous analysis of the age distribution of the terrestrial impact craters with known and precise formation age (e.g. Schmieder and Kring, 2020; Mazrouei et al., 2019) show an over-representativity of Neogene and Quaternary (0 – 23 Ma) small craters and a paucity of both small and large Precambrian (> 541 Ma) impact craters compared to other periods. The lack of large craters in the terrestrial record from the end of the Cambrian period is commonly attributed to intense erosion rates (e.g. Kenkmann and Artemieva, 2021) related to global glaciation events such as the Snowball Earth (635 – 720 Ma, e.g. Hoffman, 1998). Considering these sources of bias, we chose to restrict our analysis to craters located on stable continental landmasses (i.e. shields and platforms) (Geological Survey of Canada, 1995), for a period of time ranging from 23 Ma — 541 Ma, with uncertainties on ages lower than 10 Ma. The data and ages are extracted from recent compilation by Schmieder and Kring (2020). The extracted set of impact structures contains 45 objects ranging between 2 and 100 km (see Figure 1.a and supplementary Table A.2). The crater age range selected here (23 Ma — 541 Ma) sets the period of time that will be used for comparison with the Moon and Mars.

## 2.2. *The lunar impact crater population*

A few studies reported sets of large crater ages on the Moon including (but not limited to) Baldwin (1985); McEwen et al. (1993); Terada et al. (2020); Mazrouei et al. (2019) and Kirchoff et al. (2021). With 111 rocky craters larger than 10 km in diameter and younger than 1 Ga, Mazrouei et al. (2019) work is, up to now, the most complete set of lunar crater ages established for this size and age range. Model ages are obtained from indirect observations of associated rock abundance in lunar impact ejecta as measured with the Diviner thermal radiometer on NASA's

Lunar Reconnaissance Orbiter (LRO) (Ghent et al., 2014). A power law relationship linking rock abundance and crater age has been determined using nine anchor points corresponding to nine impact craters with ages determined by crater counts (Ghent et al., 2014) and references therein). Thus, the crater age dataset from Mazrouei et al. (2019) is dependent on crater counts and on the time-calibration function. This data set is submitted to the same interpretation restrictions than the martian crater age dataset build here (see Introduction). It is not an independent constraint for the impact flux but may be considered to discuss fluctuation of large versus small impactor population (decoupling with respect to size). In order to be conservative with the period investigated in the present study, we chose to limit this crater population to those younger than 600 Ma (see section 2.3.3), i.e. 91 craters > 10 km. Their locations are shown on Figure 1.b.

## 2.3. Martian impact craters age derivation

### 2.3.1. Crater selection and ejecta mapping

In order to robustly estimate the age of an impact event by crater counts, an impact crater should exhibit a sufficiently large number of small craters superposed on its ejecta blanket ( about 50 small craters), thus implying a large extent of the continuous ejecta deposits, and consequently a large crater diameter. The ejecta layers of 20 km diameter craters are amenable to dating. We limit our selection to the low-latitude zone between $35°N$ and $35°S$ where obliquity-driven climate variations and resulting small crater degradation and obliteration are reduced (e.g. Kreslavsky and Head, 2018); as well as to only Noachian and Hesperian regions according to Tanaka et al., (2014) to avoid any resurfacing bias in the age derivation and an over-representation of young impact craters.

We select impact craters ≥ 20 km within the 35° latitudinal band using the Robbins and Hynek (2012) database, recently revised by Lagain et al. (2021a), thus resulting in 6,554 craters. Among those entries, we discard craters devoid of an ejecta blanket, and select those with a preservation state ≥ 2 (i.e. pristine craters or presenting some degradation features without significant modification in the cavity, rims or ejecta) according to the database. All craters selected were then manually checked to eliminate those exhibiting an ejecta blanket particularly affected by tectonic, volcanic, subsequent large impacts or other geological processes, potentially compromising the retention of small craters since their formation and thus biasing the model age derivation. The final set contains 521 impact craters (see Figure 1.c).

We mapped the ejecta blanket of the 521 craters considered here using the THEMIS Day and Night IR mosaics (Edwards et al., 2011), which offer better contrast between ejecta blanket and the surrounding terrain, and limited our mapping to the CTX global mosaic coverage (Dickson et al., 2018). In order to avoid potential resurfacing of sloping rims, we removed the main impact structure from our mapping using a circular buffer area whose radius is equal to 1.2 crater radii from the centroid of the large crater (Figure 2.a).

*2.3.2. Semi-automatic model age derivation of large impact craters*

The dating of impact craters > 20 km in diameter requires the counting of numerous small craters, > 100 m superposed on the ejecta blanket. Crater counts for this size range on 521 ejecta blankets would require a too fastidious workload for only one crater expert or even a team of counters trained for this task. For instance, if all craters formed 1 Ga ago and considering a 2,500 km$^2$ counting area for each ejecta blanket, ~8,000 craters > 500 m in diameter, ~180,000 > 200 m and more than 1.5 million larger than 100 m are expected to be counted. The automation of the process is therefore a prerequisite for this approach. For this, we use a Crater Detection Algorithm (CDA), a Convolutional Neural Network developed by Benedix et al. (2020), retrained to the detection of small impact craters on high-resolution imagery covering the surface of Mars (Lagain et al., 2021b; Lagain et al., 2021c). Applied over the CTX global mosaic (Dickson et al., 2018), this method allows the derivation of consistent model ages of large martian impact craters when compared with independent manual count results if secondary impact craters are removed from the counting. This is achieved using an Automatic Secondary Crater Identification algorithm (ASCI, see Lagain et al. (2021b) for a full description of this method and validation). In brief, this tool is a cluster analysis algorithm that compares the distribution of the detected craters population on the dated area with a set of randomly distributed crater populations with similar number of craters over the same area. It then excludes craters in clusters and chains with a criteria distance based on their closest neighbors, i.e. craters are excluded if their formation is unlikely the result of a random stochastic process (meteoritic bombardment). Thus, they are considered as secondary craters in a statistical point of view.

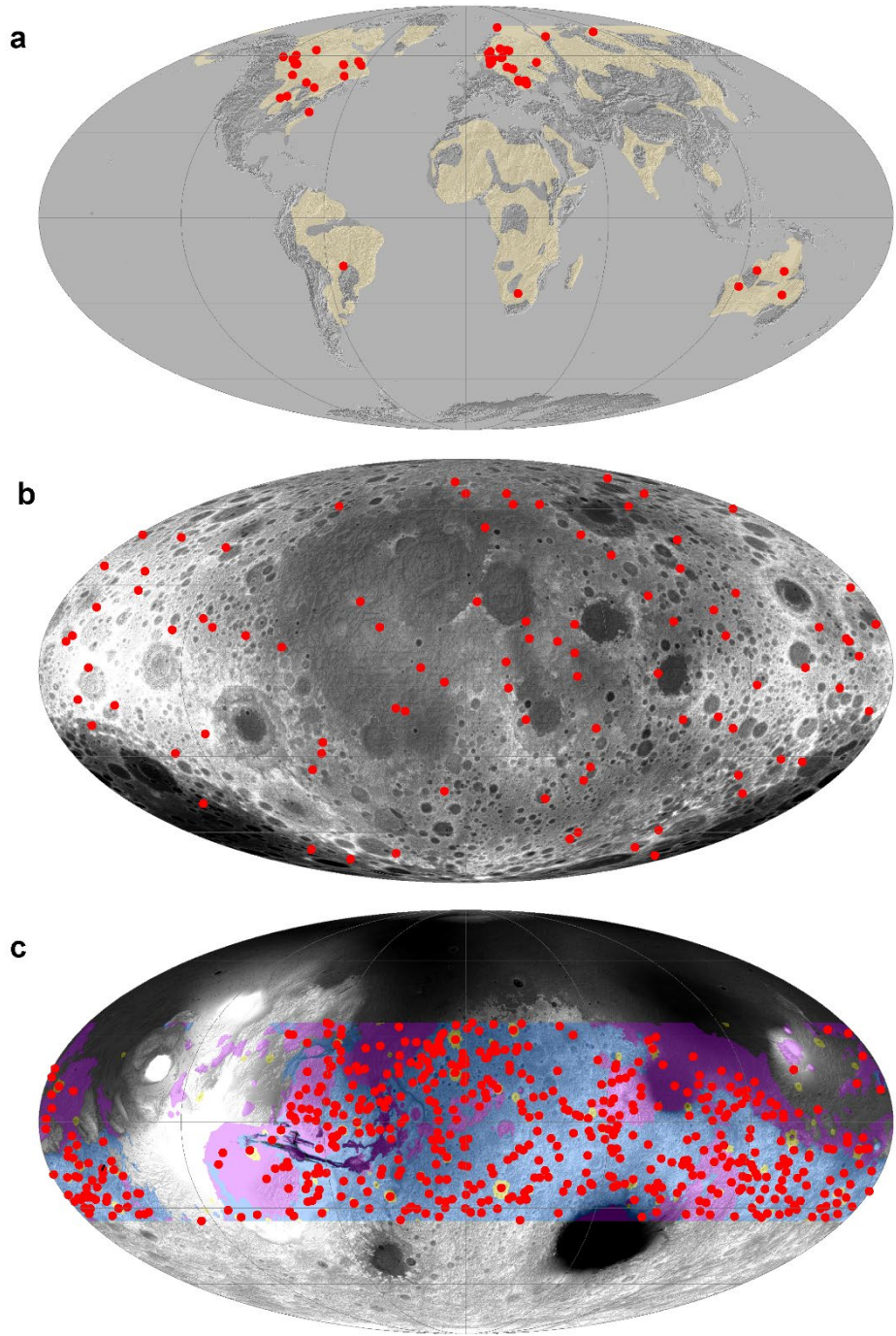

Figure 1: Location of impact craters considered in this study (red dots). (a) the Earth: All selected craters located on cratonic areas (beige areas, Geological Survey of Canada (1995)). Quaternary, Neogene and Precambrian craters as well as craters whose age errors are larger than 10 Ma are excluded (Schmieder and Kring, 2020), N = 45. The global terrain elevation data

(GMTED2010) is used as a background. (b) the Moon: All selected craters larger than 10 km in diameter and younger than 600 Ma old reported in Mazrouei et al. (2019), N = 91. The global topography (Smith et al., 2010) is used as a background. (c) Mars: All craters are >20 km in diameter and located on Noachian and Hesperian terrains (respectively blue and purple (Tanaka et al., 2014), within the ±35° latitudinal band, N = 521. The global topography (Fergason et al., 2018) is used as a background, brighter shades denote higher elevation. Mollweide projection, meridian interval: 60°, parralel interval: 30°.

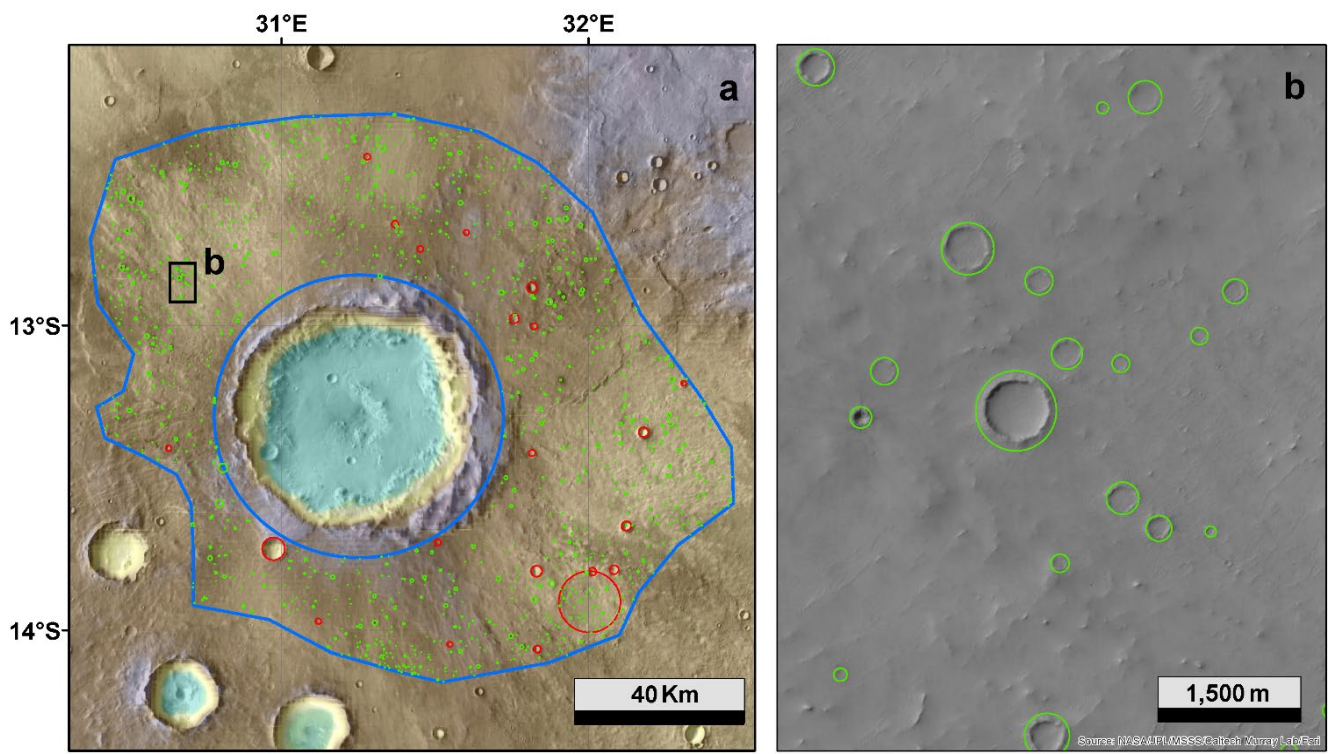

Figure 2: Crater counts on an ejecta blanket of a 40 km impact crater. (a) Ejecta blanket mapping (outlined in blue) and automatically detected craters (in green). Red circles correspond to impact craters larger than 1 km in diameter compiled in the manual crater database (Lagain et al., 2021a). Background image is the THEMIS Day-IR mosaic (Edwards et al., 2011) and colors corresponds to the elevation given by MGSMOLA- MEX HRSC Blended DEM Global 200 m v2

(Fergason et al., 2018). (b) Close-up of a part of the ejecta blanket showing detected craters over the CTX Global mosaic (Dickson et al., 2018). The diameters of the green circles have been enlarged by 15% with respect to the measured rim-to-rim diameter for better visualization of the crater rim.

Our CDA returns 1,226,387 detections larger than 100 m in diameter superposed on the ejecta blankets of the crater set. A sample of those detection is shown in green on Figure 2. Individual crater population are processed through ASCI and the remaining likely primary crater population (835,408) is used to derive their model age. For this, we use CraterStats II (Michael and Neukum, 2010) and the chronology system from Hartmann (2005), where each CSFD have been fitted with an isochrone using the Poisson analysis technique (Michael et al., 2016) to avoid any size-binning bias in the derived age. This is done using the differential form of the CSFD, thus allowing to distinguish potential resurfacing events affecting the slope of the CSFD (Figure 3). Results are presented in supplementary Table A.1. Note that the average size of small craters used to derive model ages (turn off diameter) is about 180 m, which corresponds to a ~5 m impactor Collins et al. (2005).

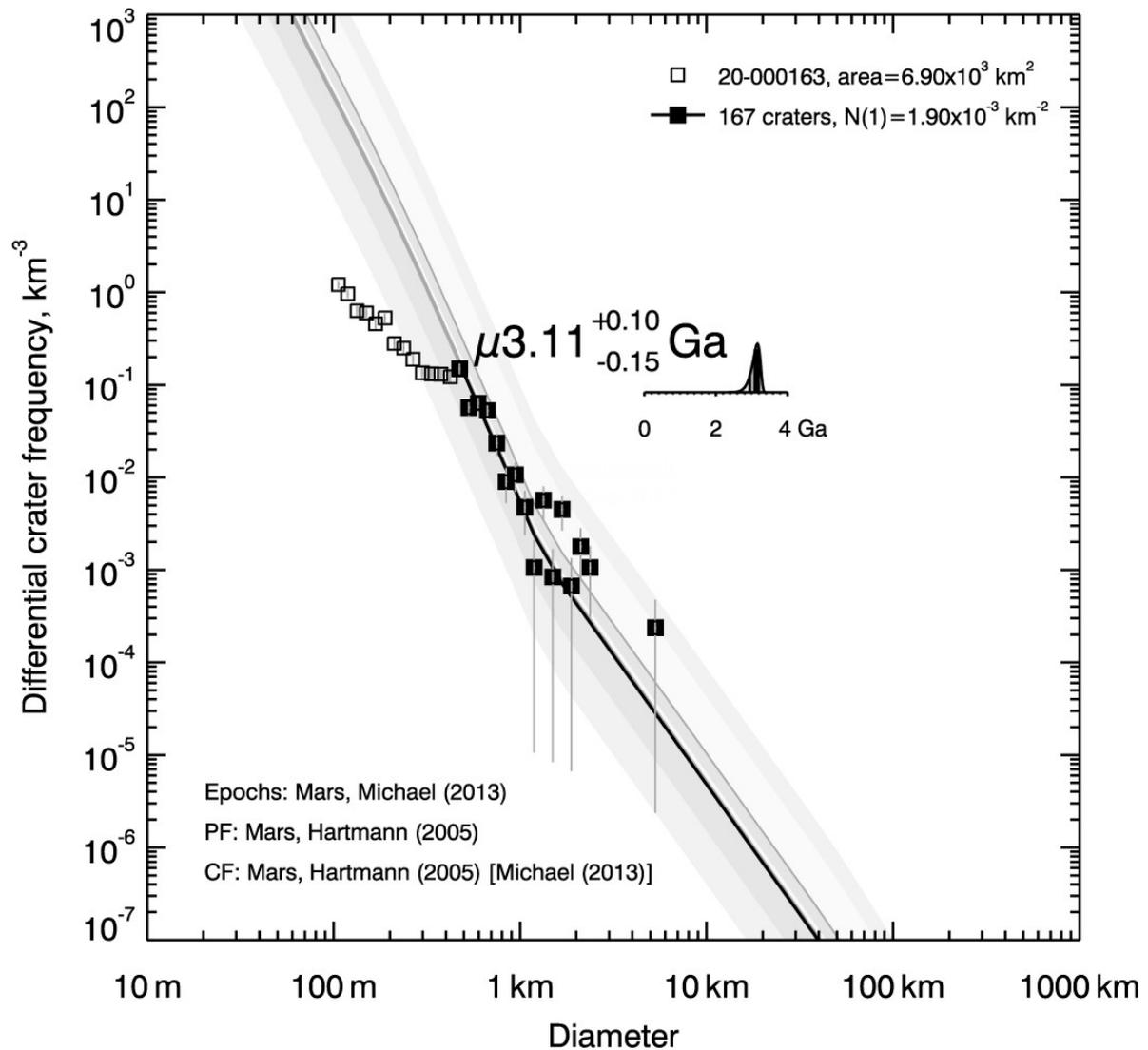

Figure 3: Model age derivation of the crater formation presented on Figure 2. The higher cratering density for some diameter bins in the 1 km - 5 km size range most likely reflects large impact craters overprinted by the ejecta blanket. The portion of the CSFD fitted here is composed by 167 craters, dominated by sizes between 300 m and 1 km. The inclusion of older craters (i.e. overlapped by the ejecta blanket) in the counts is therefore negligible in the age derivation.

*2.3.3. Completeness of the martian cratering record and age frequency distribution*

The oldest crater found in this study is ~3.8 Ga old, thus corresponding to the beginning of the Hesperian period. One can argue that the crater set considered here is not representative of the entire cratering history since ~3.8 Ga ago until present time for D ≥ 20 km on the areas investigated here due to the source of biases mentioned in section 2.3.1. We therefore determined additional constraints to build a complete crater set accumulated on the martian surface since a given period of time.

We investigate the CSFD of the large crater population and compared it against Hartmann (2005) isochrones. We selected a crater population from our set whose age falls between 0 and T, where T is the model age threshold below which all craters younger than T constitute a CSFD fitting the isochrone for the same T value. Therefore, this crater population is considered as complete for the size range considered here and located on Hesperian and Noachian terrains. It then can be used in the investigation of potential fluctuation of the impact cratering rate. As shown on Figure 4 isochrones fitting for craters older than 1 Ga shows a large discrepancy with the obtained model age, mostly due to the lack of craters > 30 – 40 km within those populations. For craters younger than 150, 300 and 600 Ma old (respectively 13, 32 and 49 craters), the obtained model age is consistent with the crater population considered (top row of Figure 4). The model age threshold from which the crater population dated here can be considered as incomplete is therefore 600 Ma: the crater population is found to be complete for craters younger than 600 Ma.

Thus, the age frequency distribution of craters younger than 600 Ma can be analyzed to investigate potential fluctuation of the production function (see supplementary Table A.1) and which could be then discussed in lights of the lunar and terrestrial data.

We characterize uncertainties in age determination using the probability density function

(PDF) associated with each age (Michael et al., 2016). Some of the ages derived in this study have significant uncertainties (> 50% of the age itself). Moreover, due to the nature of the chronology model used to derive these ages, model age uncertainties are asymmetric with larger negative uncertainties than positive. In order to consider the impact craters whose PDF intersect the completeness threshold, we compute the cratering rate from all dated craters and analyzed only the component between -600 Ma and present time. Each PDF has been normalized to unit probability by integrating the distribution over time in steps of 1 Ma to assure that each crater has an equal influence on the final impact cratering rate calculation. We then calculated the sum of all normalized PDFs. This gives us the cratering rate shown on Figure 5.

Although incomplete before 600 Ma, the impact crater flux shown on Figure 5.a shows a significant increase from -3 Ga, corresponding approximately to the end of the Hesperian period and associated with the tail-end of a protracted period of impact events in the Solar System (Bottke et al., 2012, 2015; Morbidelli et al., 2012, 2018; Fritz et al., 2014). We also note that 80% of craters > 40 km are older than 3 Ga while only 50% of craters > 20 km and < 40 km are formed over the same period of time. If we assume that the incompleteness of the cratering record due to erosion processes affects the crater population in the same way for all crater sizes, this result might reflect the modification of the size-frequency distribution of impactors at that time, in accordance with orbital and collisional simulations (Morbidelli et al., 2012, 2018; Bottke et al., 2012, 2015). However, larger craters exhibit larger and thicker ejecta blanket and can be dated using larger superposed craters, more resistant to the erosion. Thus, the incompleteness of the record might not be crater size-independent and this result could only reflect a sampling bias when the selection of impact craters has been made (section 2.1), i.e., a lack of sampling for craters

< 40 km in diameter and older than 3 Ga. Therefore, our data cannot be used to conclude on the size-dependency of the early evolution of the cratering flux.

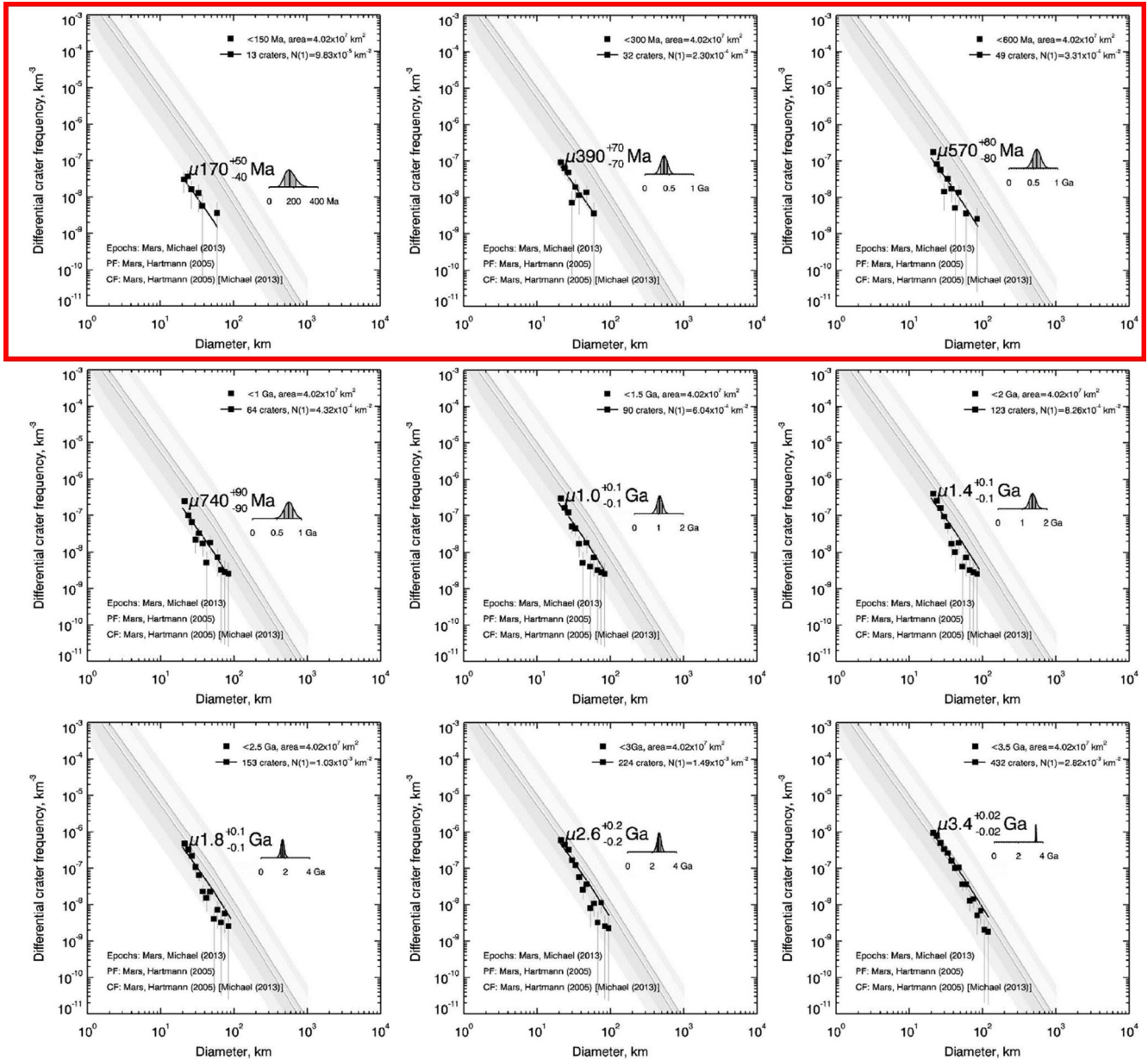

Figure 4: CSFDs and model age fitting for different dated crater population younger than T. Isochrones fitting are in good agreement with CSFDs of craters younger than 600 Ma (top row).

According to our model age derivations, only 49 out of 521 dated impact craters are younger than 600 Ma and constitute the complete portion of the cratering record on martian areas investigated here. The relative impact flux derived from those craters model age shows an abrupt decrease from -300 Ma (Figure 5.b) with a flux 1.9 times higher in the 0 – 300 Ma period compared to the 300 – 600 Ma period. This ratio rises up to 2 if both the first and the last 50 Ma age bins of the 0 – 600 Ma period are removed from the calculation. We discuss the statistical significance of this result in the next section.

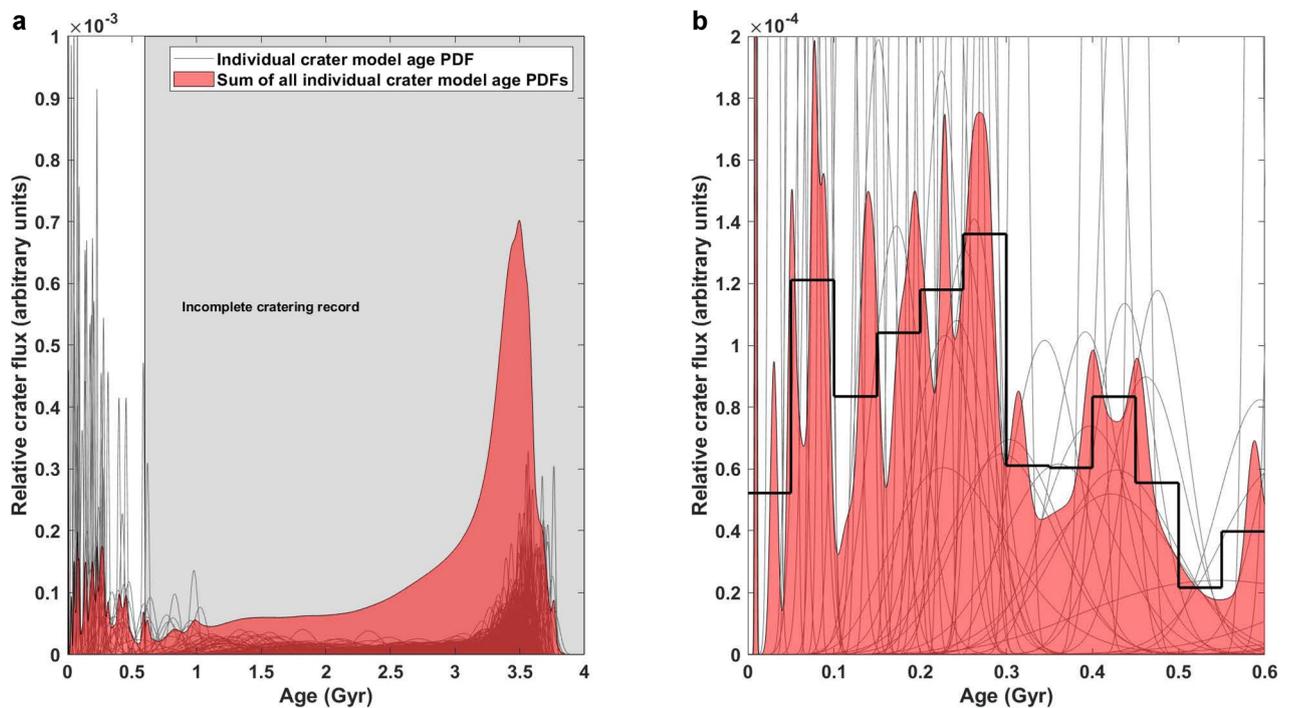

Figure 5: Impact cratering flux inferred from crater model age derivation. (a) Grey curves correspond to individual crater age PDF normalized to unit probability and the red curve corresponds to the relative cratering flux obtained from the sum of all normalized PDFs. (b) Close-up of panel (a) showing the relativecrater flux over the 0 – 600 Ma age range. Black steps correspond to the relative crater flux integrated by 50 Ma steps.

## 3. Analysis of impact crater age distributions

In this section, we assess the randomness of the crater age distribution of the three considered populations on the Earth, Moon, and Mars. Note that the crater "age" presented on Figures 5, 6b and 6c is a measure of the accumulated small crater density. It is often expressed in terms of "model age" for illustration and convenience only. It can be considered as a proxy of the true age only under the assumption that the cratering rate of both large and small craters is constant (see Introduction).

If the small crater impact rate was strictly proportional, coupled, to the large crater impact rate, then the ages of craters would be a random sample of a uniform distribution. We test this formal statistical hypothesis with a Kolmogorov – Smirnov test applied on the crater populations of the 3 bodies considered here. The Kolmogorov-Smirnov test (e.g., Daniel, 1990) is a common non-parametric statistical test widely used to compare a sample with a reference probability distribution. It is based on the maximum difference between the observed and reference cumulative frequency distributions. It is sensitive to general, long-term changes, and thus provides statistical evidence of this type of impact rate change. However, systematic age biases (for example, preferential obliteration of small craters for older ages) should be considered when interpreting non-random results. Furthermore, this test is not sensitive to sharp short-term rate peaks. Therefore, we develop the following Monte-Carlo-style statistical test especially sensitive to short-term rate peaks. We sort the craters according to their age. For each set of $L$ consecutive craters, we calculate the mean impact rate as $L$ divided by the age difference between the oldest and the youngest craters in the set. We assign this rate to the median age within the set. Then we found the maximum rate over all sets of L consecutive craters. This gives us the peak rate, our test statistics. Then we generate a set of random ages uniformly distributed and calculate the peak rate for this random age set. We repeated this for $5 \times 10^5$ random age sets, which gives us the peak rate frequency distribution for uniformly distributed ages, which we use to obtain the p-value for our real age set. It is commonly admitted that a small p-value (< 0.05) is needed to reject non-randomness.

The Kolmogorov – Smirnov test applied on the 45 terrestrial impact craters considered here yields the p-value of 0.09. This high p-value allows rejecting long term fluctuation of the cratering rate. The peak rate test is the most sensitive for $L = 9$ craters and yields p-value of 0.002. This value is sufficiently low to seriously consider non-randomness of terrestrial impact formation. Figure 6.a shows the history of the terrestrial impact rate calculated with $L = 9$ craters. It is seen that the significantly low p-value is caused by the proposed Ordovician peak at ~450 Ma discussed in the next section.

Model ages of lunar craters > 10 km and younger than 600 Ma reported by Mazrouei et al. (2019) yields a rather low p-value of 0.009 by the Kolmogorov – Smirnov test. This reflects the long-term rate change at the Mesozoic / Paleozoic boundary reported and discussed by Mazrouei et al. (2019). For the peak rate test, the lowest p-value of 0.08 is achieved for $L = 13$ craters (Figure 6.b), reached at a very young age of ~40 Ma. Although Neogoene and Quaternary craters have been discarded in the terrestrial crater population considered here, this late peak most likely reflects a preservation bias, leading to an over-representation of very young craters. Thus, this peak should not be considered as statistically significant. The data of Mazrouei et al. (2019) do not show any prominent short rate increase. As there is no reason to think that the relative cratering evolution of the Moon and the Earth is different, this results raised questions in the using of both population to infer impact flux variation, discussed in the next section. Accounting for the martian crater model age population (supplementary Table A.1), the Kolmogorov - Smirnov test yields the p-value of 0.11, rather high value, which would reject age randomness only at "one-sigma" level. Figure 6.c shows the history of impact rate calculated with $L$ = 11 craters; a peak observed at ~250 Ma on Figure 5.b is clearly seen. The lowest p-value of 0.08 is reached for $L$ = 11 craters. This means that in ~8% of the purely random age sets the peak rate exceeded the observed one. This is lower than the $2\sigma$ confidence level commonly admitted to define a significant statistical signal. Since we specially adjusted $L$ to obtain the highest test sensitivity, we need to be conservative in our choice of the significance level that would reject randomness. The obtained p-value of 0.08 is certainly not conservative enough, and we have to conclude that the observed peak in impact rate at ~250 Ma is not statistically significant.

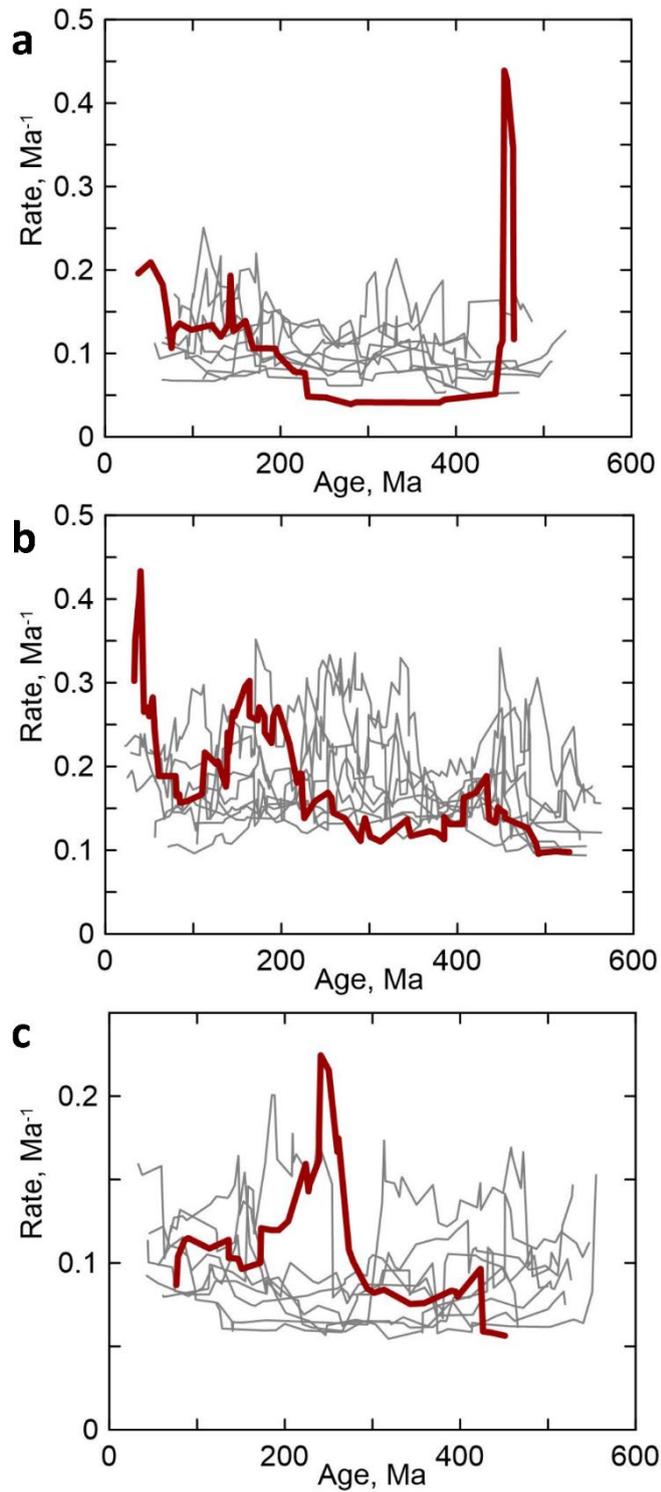

Figure 6: Impact rate calculated from measured impact crater ages (bold curve) compared to that of eight examples with random ages uniformly distributed (thin curves). (a) the Earth, calculated for 45 terrestrialcraters on the cratons with $L$ = 9 craters and age interval from 23

Ma to 541 Ma. (b) the Moon, calculated for 91 lunar craters > 10 km and younger than 600 Ma, $L$ =13 craters. (c) Mars, calculated for 49 craters younger than 600 Ma, $L$ = 11 craters window.

## 4. Discussion

### 4.1. *The terrestrial record*

The proximity of Mars to the main belt, as well as the Earth-Moon distance both exclude the possibility that one of these three bodies experienced a cratering spike in their geological history whether the others did not. Although the lunar cratering rate inferred by rock abundance from Mazrouei et al. (2019) presents a significant increase in recent time, its past evolution over older periods of the Phanerozoic is not characterized by either a short or long- term rise (Figure 6.b). Not any significant spike has been deduced from our measurements on Mars. The absence of such signal in the lunar and martian cratering record raises questions about the qualitative increase observed on the Earth.

The terrestrial geological record indicates a brief period of intense increase of dust flux (three to four order of magnitudes above the background) following the L-Chondrite Parent Body (LCPB) break-up event during the Ordovician period, 466 Ma ago (e.g., Schmieder and Kring, 2020). Even in present time, this asteroid break-up is the source of about one-third of all meteorites recovered on Earth. The impact craters age frequency distribution also exhibits an anomaly for the same period despite preservation bias of impact structures on Earth discussed in section 2.3, with one order of magnitude higher number of craters in the 440 – 470 Ma range compared to other Phanerozoic periods, i.e., the last ~541 Ma (e.g., Schmieder and Kring, 2020). However, this anomaly in the cratering record is dominated by craters smaller than 10 km in diameter (corresponding to asteroids smaller than about 1 km in diameter). Only 2 out of 10 craters in the 445 – 466 Ma age range are larger than 10 km, corresponding to East Clearwater and State Islands. Although a cluster of Ordovician impact craters does exist on Earth (e.g. Schmieder and Kring, 2020; Kenkmann and Artemieva, 2021), a sharp increase in the absolute impact flux at that time due to an underlying punctuated phenomenon (i.e. an asteroid breakup, e.g. Alwmark and Schmitz, 2007) is questionable.

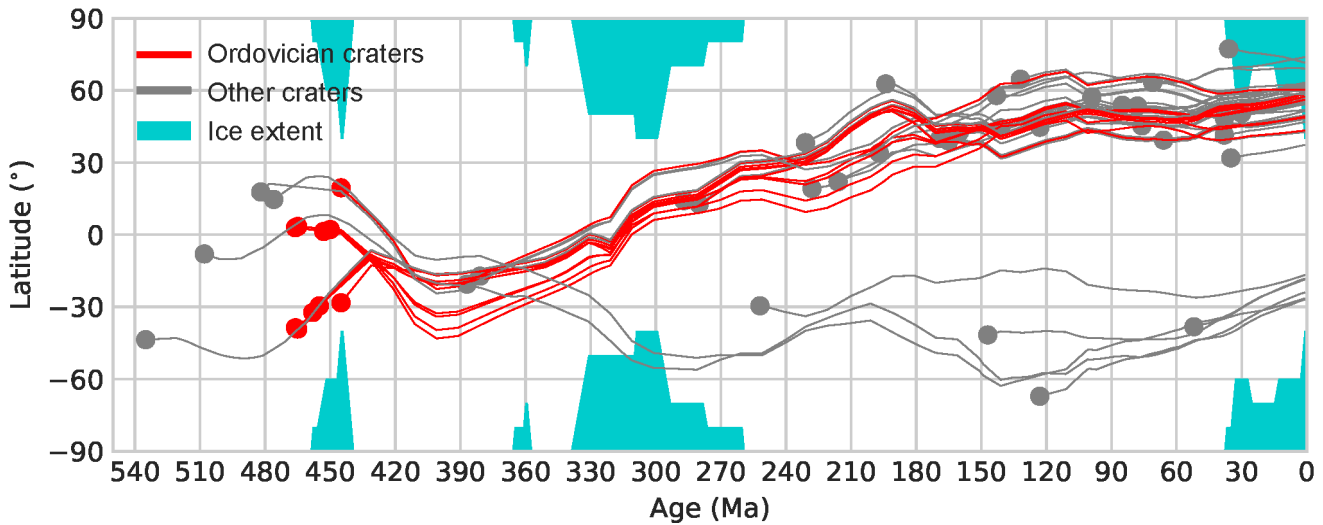

Figure 7: Latitudinal reconstruction of the path followed by Ordovicin (red lines) and other Paleozoic impact craters (grey lines) presented in section 2.3 since their formation (dots). Crater ages are from Schmieder and Kring (2020), their position evolution is based on full plate tectonics model (Merdith et al., 2021; Macdonald et al., 2019) using GPlates software (Müller et al., 2018), and polar ice cap extent (cyan areas) is from Pohl et al. (2016).

Impact crater burial and exhumation are dependent of the polar ice cap extent, sedimentation rate, and landmasses distribution. Using plate tectonics reconstruction software (GPlates, Müller et al., 2018, licensed for distribution under the General Public License), updated model reconstructing paleogeographic features through the Phanerozoic era (Merdith et al., 2021) and impact event ages, we infer the formation location of the 45 craters presented in section 2.3. Figure 7 presents the latitudes of the formation of those craters and the evolution of their position from plate tectonics model (Merdith et al., 2021). Extent evolution of the polar ice caps is also represented in blue (Macdonald et al., 2019).

While impact craters < 200 Ma old considered here were formed at mid-latitudes (~30° of latitude) and beyond, the craters constituting the Ordovician cluster were formed at tropical latitudes (Figure 7). Numerical climate models and carbon isotope measurements suggests that atmospheric levels of carbon dioxide during the Ordovician period were 14–16 times higher than today, mostly driven by widespread volcanic activity (Finnegan et al., 2012). The extensive flooding of continents due to high sea levels (up to 200 m compared to present level), combined with the lack of widespread vegetation on land, would have suppressed the weathering of silicate rocks, leading to a positive atmospheric $CO_2$ budget over most of the Ordovician period (Munnecke et al., 2010). This has caused a global greenhouse effect, and thus, a rise of temperatures from the equator to the poles (Cocks and Torsvik, 2020). Clear evidence of warm temperatures in the tropics can be seen in the extensive Ordovician limestone deposits (e.g. Cocks and Torsvik, 2020). Extended epicontinental shallow seas, marked by intense carbonated platforms development, were then drained during the late Ordovician period due to the drop in sea level caused by the rapid drop of global temperatures, leading to the Hirnantian ice age (Villas et al., 2002) (Figure 7). Out of the 10 impact craters falling into the 445-466 Ma age range considered here, 3 were formed in a shallow marine condition and 6 in sedimentary layers (e.g.

Kenkmann, 2021).

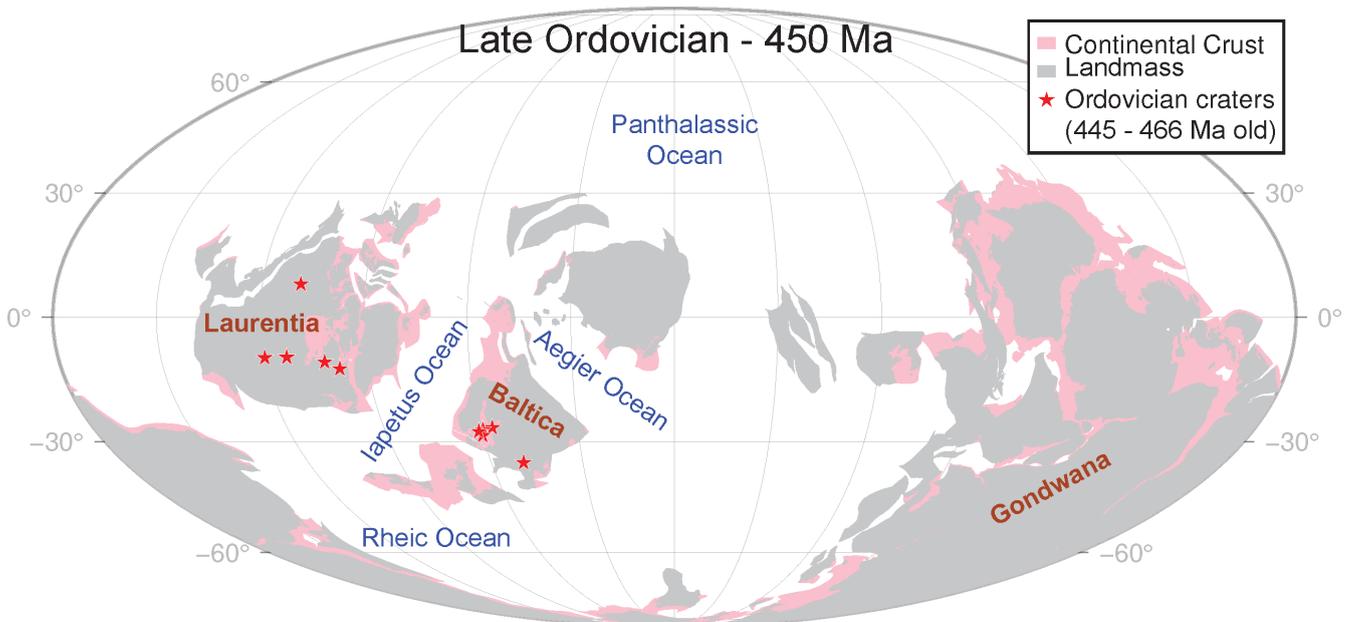

Figure 8: Configuration of continental landmass during 450 Ma ago, using full plate reconstruction models from Merdith et al. (2021) and Macdonald et al. (2019) using GPlates software (Müller et al., 2018), showing the approximate formation location of Ordovician craters. Present-day coastlines are shown in grey. Note the high concentration of impact craters on the western part of Baltica (corresponding to the present Scandinavia).

The environmental conditions marking the Ordovician period have thus most likely led to a better preservation of impact craters formed in the tropical region followed by a rapid burial by sedimentary layers, that were later exhumed during glacial periods, where continental masses were located at higher latitudes. Moreover, impact craters are often more precisely dated, when their association with well-characterized sedimentary strata is known (Schmieder and Kring, 2020; Jourdan et al., 2009), an effect that contributes to the high concentration of Ordovician impact craters in Scandinavia (western part of Baltica on Figure 8) (e.g. Schmieder and Kring, 2020). The particular environmental condition at the Ordovician most likely contributes to the apparent cratering spike observed here and thus might not be associated with the LCPB asteroid break-up. By extension, we can hypothesize that the distribution of landmasses and environmental conditions during post-Ordovician periods most likely did not favor the preservation of younger impact structures, thus leading to an incompleteness of the crater record.

*4.2. Asteroid break-up and impact flux variation mechanism*

Orbital dynamic mechanisms have been invoked to explain potential fluctuation of the impact flux. The Yarkovsky effect (anisotropic heat emission that change the semi-major axis of asteroids) efficiency in feeding main resonances of the inner and central main belt such as the $v_6$, 3:1 and 5:2 in 0.1 m − 10 km objects is non-negligible and has been suggested to explain potential long-term fluctuation in the impact cratering flux observed on the Moon (Mazrouei et al., 2019; Kirchoff et al., 2021; Terada et al., 2020). Nevertheless, the size-dependency of this process impose a decoupling in the cratering flux between impactors of different size because the small fragments shower following a considered asteroid break-up would happen earlier and will be shorter than the bigger ones. However, small debris are thought to be grinded through collisional cascades before reaching escape routes towards planets-crossing orbits. In other words, size-frequency distributions of 0.1 – 100 m asteroids generated after a catastrophic disruption follows that of the main belt preexisting one within a couple of million years (Bottke et al., 2015) and therefore do not dominate the background of impactors even in the case of a break-up occurring at the immediate vicinity of unstable orbits (i.e., resonances). Therefore, sharp increase in the flux of large asteroids is possible if a massive disruption occurs really close to a delivery resonance, thus reducing the transit time through Yarkovsky effect. An asteroid family orbiting further away from a delivery resonance would either be too small to create a significant peak of material;

and/or the spike would be long-lived if it was large enough to send a recordable amount of material to Mars-crossing orbits or in Near-Earth space.

The Baptistina and Flora asteroids family, formed at proximity of resonances following catastrophic disruptions 140 Ma and 500 Ma ago respectively, have been proposed as the source of a recent increase in the member of the Near-Earth Objects population (Bottke et al., 2007; Nesvorný et al., 2007; Marchi et al., 2009) for sub-kilometer asteroids (Hartmann et al., 2007). The small debris following such breakups escape the main belt too quickly to be grinded by collisional cascades. Those events may be the most important process that produce a non-stationary stochastic flux responsible of temporal varying impact cratering rate. However, short and intense spike might produce the same increase in the cumulative cratering record than a longer and gentler spike. The detectability of such increase depends directly on the spike integral, but also, in the frame of our technique, on how much decoupled the flux of small and large impactors is.

As mentioned above, the potential decoupling depends on the size of craters/impactors considered. Only meteorite and micrometeorite size debris from the formation of these dynamical families might have significantly affected the recent impact flux on terrestrial planets (Hartmann et al. , 2007; Terfelt and Schmitz, 2021; Schmitz et al., 2019) and the cratering rate on the Moon and Mars (small debris being filtered by the Earth's atmosphere (Bland, 2005)).

### 4.3. A late increasing in the impact flux?

The crater age frequency distribution on Earth and the Moon appears to increase to younger periods, i.e. after 290 Ma, and has been interpreted as a rise in the impact flux relative to the 300 – 650 Ma period (Mazrouei et al., 2019) rather than a preservation bias. However,

Hergarten et al. (2019) have shown that the age-frequency distribution observed on Earth can be reproduced by a constant crater production, associated with an erosion constant in time, but variable in space. They also proposed an alternative model linking the rock abundance and lunar crater ages used by Mazrouei et al. (2019) to determine the ages of a lunar crater set. Rather than assuming a power-law relationship derived from nine craters of known ages (Ghent et al., 2014), the authors show that the distribution can be approximated by an exponential function, producing a fit of similar quality than that proposed by Ghent et al. (2014). Using this relationship, the increase in impact flux inferred from Mazrouei et al. (2019) vanishes. Thus, the apparent increase in the lunar and terrestrial impact flux is consistent with a constant impact flux under a differential terrestrial erosion rate and a different lunar rock abundance temporal evolution model.

Our results show a discrepancy in the age frequency distribution of impact craters considered here for Mars, the Earth and the Moon. Following impact flux variation mechanisms discussed in section 4.2, and potential issues in the calibration method of lunar data presented above, and crater preservation biases on Earth developed in section 4.1, differences in the age-frequency distributions for the three bodies we investigated can be interpreted in three ways:

1. the absence of a statistically significant decoupling between the flux of small and large asteroids on Mars does not preclude its existence. In that case, uncertainties linked to the model age derivation technique on the Moon and Mars might affect the detectability of the flux decoupling. Alternatively, the impact flux might have varied under a couple regime between the small and large asteroids, with alternations of spikes and lulls periods of the same intensity for all asteroid sizes. In this case, the technique employed here is blind regarding such fluctuations.

2. the impactor size where the flux may switch from constant to non-constant is below the associated crater size considered on Mars (D > 180 m, so impactors larger than ~5 m),

3. the over-represented population of Ordovician craters on Earth is favored by preservation bias and is not linked to a significant increase of the impact flux of large asteroids. The relative late increase of the flux measured on the Moon (Mazrouei et al., 2019) is most likely due to systematic errors in the age derivation method. In that case, the impact flux on Mars and in the Earth-Moon system did not significantly change and remained relatively coupled between that of the small and large impactors over the last 600 Ma.

Considering delivery processes of impactors following an asteroid break-up, including collisional cascades, Yarkovsky effect and resonances are controlled by object size, and considering that a cratering peak of non-stochastic origin would have been visible on all bodies, the different signals analyzed here are clearly against a non-statistical fluctuation of the impact flux. Thus, the first interpretation can be ruled out. Regarding the second interpretation, the dating of large impact craters on Mars has been performed using craters > 180 m (see turn off diameter in supplementary Table A.1), thus corresponding to asteroids > 5 m (Collins et al., 2005). In comparison, craters > 20 km on Mars are formed by ~1.7 km asteroids and more. This range of impactor size covers largely the size of large asteroids having formed Ordovician craters on Earth and the lunar crater population analyzed here, and for which a cratering spike and a decoupling in the flux of different size of impactors is unlikely over the last 600 Ma. However, our data cannot reject the possibility that the impactor flux for D < 5 m significantly varied with respect to that of kilometric asteroids (second hypothesis).

Therefore, we favor the third scenario and argue that relative temporal fluctuation of the impact flux in the inner Solar System between asteroids >2 km and 5 m − 100 m is relatively limited, statistically insignificant over the last ~600 Ma. This is consistent with with the traditional model for delivering asteroids to planet-crossing. The recent dating of new lunar samples from the Change5 mission (Figure 4 in Che et al., 2021) and the associated cumulative crater densities measured in the immediate vicinity of the landing site (Qiao et al., 2021) support these concluding as these data agree with current chronology models assuming a constant and coupled impact flux over the last ~3 Ga.

## 5. Conclusion

The asteroid impact flux is commonly assumed to be constant in the inner Solar System over the last ~3 Ga. The crater production rate for all sizes on terrestrial surfaces is also assumed to be in a steady state. Those assumptions are widely used by the community to date planetary surfaces using crater counts. However, they are challenged by several studies analysing the crater age distribution on the Moon and the Earth (e.g. Terada et al., 2020; Kirchoff et al., 2021; Mazrouei et al., 2019) arguing for the existence of periods of lull and spikes in the impact flux, thus modifying the impact cratering record.

In this study, we used a formal statistical analysis to assess any fluctuation of the crater-size distribution with time on Mars, and discuss the impact flux in the inner Solar System by integrating lunar data and independent constraint from the recent terrestrial crater record.

Our statistical tests return non-statistically significant fluctuation of the martian cratering rate. The impact cratering flux on Earth shows a cratering spike during the Ordovician, about 470 Ma ago. Using plate tectonic models, we estimate the location of the impact events we consider here. We found that craters constituting the Ordovician spike were formed in tropical regions. Coupled with the geological record, we argue that the Ordovician spike is the result of a preservation bias rather than a real increase in the asteroid impact flux. The lunar crater age distribution, obtained through the relationship between the rock abundance in the ejecta blanket and the age of large craters, shows a statistically significant late increase over the last ~300 Ma. Since other calibration methods used to date lunar impact events are consistent with a constant crater production rate, we argue that this late increase is due to uncertain calibration method.

This rejects both cratering spikes and late increasing of the impact flux in the Earth-Moon system, the latter being proposed by Mazrouei et al. (2019). This also raise issues about the lunar crater dating method accuracy developed by Ghent et al. (2014), as detailed by Hergarten et al. (2019), and subsequent broader implications such as in Černok et al. (2021). According to the latter, the 0.5 Ga event seen in one of the Apollo 17 sample (inferred from disturbance of U-Pb chronometer) is attributed to a distant impact that has formed Dawes crater (dated at ~0.5 Ga using Ghent et al. (2014) method by Mazrouei et al., 2019)) and that has transported materials on the sampling site located on .

Nevertheless, the existence of spikes in the impact flux for meteorite-size impactors, possibly linked to the LCPB event (Terfelt and Schmitz, 2021; Schmitz et al., 2019), requires to be tested on the Moon. To that end, the using of semi-automatic dating techniques such as the one presented in this study is a pre-requisite to access to the model age of a large statistical sample of Copernician craters model age using small crater count.

**Acknowledgments**

This work was supported by the resources provided by the Pawsey Supercomputing Centre with funding from the Australian Government, Curtin University, and the Government of Western Australia. A. Lagain and G.K. Benedix are funded by the Australia Research Council (DP170102972 and FT170100024). This work was also partly supported by NASA grant NNX16AQ06G to M. Kreslavsky. We are grateful the Murray Lab for the building and the release of the CTX Global mosaic. We thank the Curtin Hub for Immersive Visualization and eResearch (HIVE) for their help in the visualization of our crater detection having allowed the best use of our algorithm. The data used are listed in the references, in Appendix A, and the details of martian craters age derivation and crater counts are available at https://doi.org/10.5281/zenodo.5768723.

**Appendix A. Appendix**

Table A.1: Impact craters younger than 600 Ma old. If unnamed, crater IDs are from Robbins and Hynek (2012). N(1) corresponds to the number of craters with a diameter larger than or equal to 1 km per km$^2$, derived by fitting the standard distribution (Hartmann, 2005) to the observed CSFD. The turn off diameter corresponds to the minimum diameter used to fit the CSFD with an isochron.

| Crater ID | Diameter | Latitude | Longitude | Turnoff | Model age | N(1) |
|---|---|---|---|---|---|---|

|           | (km) |       |        | diameter (m) | (Ma)  |              |
|-----------|------|-------|--------|--------------|-------|--------------|
| Mojave    | 58.0 | 7.5   | -33.0  | 100          | 9.2   | $5.37 \times 10^{-06}$ |
| Santa Fe  | 20.3 | 19.3  | -47.9  | 100          | 30.1  | $1.76 \times 10^{-05}$ |
| 16-000223 | 33.9 | -17.8 | -157.5 | 100          | 50.2  | $2.93 \times 10^{-05}$ |
| 24-000696 | 21.1 | -31.7 | -148.0 | 140          | 55.2  | $3.22 \times 10^{-05}$ |
| 16-000200 | 36.3 | -2.4  | -177.0 | 160          | 70.7  | $4.13 \times 10^{-05}$ |
| 22-000295 | 26.8 | -9.8  | 101.2  | 100          | 76.4  | $4.46 \times 10^{-05}$ |
| 12-000475 | 25.5 | 17.7  | 39.7   | 170          | 78.2  | $4.57 \times 10^{-05}$ |
| 04-000178 | 22.2 | 30.4  | -57.6  | 140          | 85.7  | $5.01 \times 10^{-05}$ |
| 23-000425 | 23.2 | -26.9 | 139.8  | 100          | 89.9  | $5.25 \times 10^{-05}$ |
| 11-000353 | 22.7 | 21.5  | -1.5   | 160          | 114.0 | $6.64 \times 10^{-05}$ |
| 19-000358 | 24.8 | -3.7  | -11.8  | 100          | 136.0 | $7.93 \times 10^{-05}$ |
| 22-000189 | 34.8 | -18.5 | 98.7   | 140          | 136.0 | $7.92 \times 10^{-05}$ |
| 23-000431 | 22.9 | -9.3  | 169.4  | 100          | 147.0 | $8.57 \times 10^{-05}$ |
| Xainza    | 24.0 | 0.8   | -3.9   | 240          | 151.0 | $8.80 \times 10^{-05}$ |
| 12-000479 | 25.3 | 7.1   | 41.0   | 200          | 172.0 | $1.00 \times 10^{-04}$ |
| 29-000688 | 22.1 | -30.1 | 131.8  | 100          | 173.0 | $1.01 \times 10^{-04}$ |
| 08-000041 | 33.0 | 1.0   | -176.8 | 150          | 185.0 | $1.08 \times 10^{-04}$ |
| 22-000326 | 25.3 | -16.0 | 124.0  | 100          | 193.0 | $1.13 \times 10^{-04}$ |
| 18-000147 | 20.4 | -17.7 | -60.5  | 100          | 204.0 | $1.19 \times 10^{-04}$ |
| 15-000086 | 20.9 | 13.3  | 167.5  | 150          | 224.0 | $1.31 \times 10^{-04}$ |
| 18-000062 | 35.0 | -20.2 | -45.9  | 100          | 227.0 | $1.33 \times 10^{-04}$ |
| 11-000352 | 22.8 | 11.8  | -35.9  | 230          | 228.0 | $1.33 \times 10^{-04}$ |
| 11-000311 | 25.1 | 11.3  | -16.5  | 170          | 239.0 | $1.39 \times 10^{-04}$ |
| 13-000228 | 31.1 | 1.9   | 48.0   | 350          | 239.0 | $1.39 \times 10^{-04}$ |
| 10-000140 | 23.4 | 27.6  | -72.0  | 220          | 241.0 | $1.41 \times 10^{-04}$ |
| 17-000030 | 25.2 | -14.7 | -106.3 | 250          | 250.0 | $1.46 \times 10^{-04}$ |
| 14-000035 | 47.8 | 10.2  | 94.3   | 200          | 260.0 | $1.52 \times 10^{-04}$ |
| 22-000463 | 20.7 | -22.5 | 95.5   | 200          | 261.0 | $1.53 \times 10^{-04}$ |
| 11-000425 | 20.0 | 27.8  | -22.5  | 100          | 266.0 | $1.55 \times 10^{-04}$ |
| 11-000102 | 46.3 | 14.8  | -18.9  | 150          | 273.0 | $1.60 \times 10^{-04}$ |
| 11-000100 | 47.1 | 23.7  | -14.3  | 150          | 278.0 | $1.62 \times 10^{-04}$ |
| 19-000427 | 21.5 | -16.7 | -28.2  | 300          | 294.0 | $1.72 \times 10^{-04}$ |

*Table A.1 continued*

| Crater ID | Diameter (km) | Latitude | Longitude | Turn off diameter (m) | Model age (Ma) | N(1) |
|---|---|---|---|---|---|---|
| 13-000220 | 32.1 | 7.3 | 58.4 | 350 | 302.0 | $1.77 \times 10^{-04}$ |
| 11-000321 | 24.5 | 8.7 | -27.6 | 100 | 313.0 | $1.83 \times 10^{-04}$ |
| 21-000302 | 32.8 | -17.6 | 52.2 | 260 | 343.0 | $2.00 \times 10^{-04}$ |
| 18-000144 | 20.4 | -16.1 | -82.2 | 280 | 359.0 | $2.10 \times 10^{-04}$ |
| 26-000315 | 27.8 | -34.5 | -31.0 | 250 | 390.0 | $2.28 \times 10^{-04}$ |
| 10-000068 | 37.0 | 6.9 | -59.1 | 310 | 395.0 | $2.31 \times 10^{-04}$ |
| Phedra | 20.3 | 13.8 | 123.9 | 100 | 397.0 | $2.32 \times 10^{-04}$ |
| 22-000377 | 23.2 | -28.3 | 109.5 | 220 | 419.0 | $2.45 \times 10^{-04}$ |
| 12-000036 | 82.0 | 11.5 | 13.5 | 280 | 423.0 | $2.47 \times 10^{-04}$ |
| 23-000459 | 21.9 | -9.2 | 171.0 | 250 | 426.0 | $2.49 \times 10^{-04}$ |
| 18-000151 | 20.1 | -17.9 | -65.3 | 220 | 435.0 | $2.54 \times 10^{-04}$ |
| Guaymas | 20.1 | 25.7 | -45.0 | 100 | 451.0 | $2.64 \times 10^{-04}$ |
| 21-000599 | 21.2 | -19.9 | 73.3 | 220 | 460.0 | $2.68 \times 10^{-04}$ |
| Concord | 20.5 | 16.5 | -34.0 | 150 | 473.0 | $2.76 \times 10^{-04}$ |
| 13-000230 | 30.9 | 6.4 | 53.2 | 600 | 577.0 | $3.37 \times 10^{-04}$ |
| 11-000123 | 41.6 | 20.5 | -24.1 | 100 | 584.0 | $3.41 \times 10^{-04}$ |
| 23-000508 | 20.6 | -17.0 | 175.6 | 200 | 592.0 | $3.46 \times 10^{-04}$ |

Table A.2: **List of the 45 terrestrial craters Schmieder and Kring (2020) shown on Figure 1.a and paleolatitudes of their formation as presented on Figure 7.**

| Name | Diameter (km) | Latitude (present time) | Longitude (present time) | Age (Ma) | Latitude (formation time) |
|---|---|---|---|---|---|
| Logoisk | 17 | 54.2 | 27.8 | 30.0 | 50.4 |
| Chesapeake | 42.5 | 37.3 | -76.1 | 34.9 | 31.9 |
| Popigai | 100 | 71.5 | 111.0 | 35.7 | 77.2 |
| Wanapitei | 7.5 | 46.8 | -80.8 | 37.7 | 41.4 |
| Mistastin | 28 | 55.9 | -63.3 | 37.8 | 48.5 |
| Goat Paddock | 5.1 | -18.3 | 126.7 | 52.0 | -38.2 |
| Boltysh | 24 | 48.8 | 32.2 | 65.8 | 39.3 |
| Kara | 65 | 69.1 | 64.3 | 70.7 | 63.3 |
| Manson | 35 | 42.6 | -94.5 | 75.9 | 45.2 |
| Lappajärvi | 23 | 63.2 | 23.7 | 77.9 | 53.5 |

| Suvasvesi North | 3.5 | 62.7 | 28.2 | 85.0 | 53.9 |
| Deep Bay | 13 | 56.4 | -103.0 | 98.5 | 57.3 |

*Table A.2 continued*

| | | | | | |
|---|---|---|---|---|---|
| Mien | 9 | 56.4 | 14.9 | 122.4 | 44.7 |
| Tookoonooka | 55 | -27.0 | 143.0 | 122.8 | -67.1 |
| Steen River | 25 | 59.5 | -117.6 | 132.0 | 64.6 |
| Dellen | 19 | 61.9 | 16.7 | 140.8 | 44.4 |
| Mjølnir | 30 | 73.8 | 29.7 | 143.0 | 57.9 |
| Morokweng | 70 | -26.5 | 23.5 | 146.1 | -41.6 |
| Vepriaj | 8 | 55.1 | 24.6 | 160.0 | 42.8 |
| Zapadnaya | 3.2 | 49.7 | 29.0 | 165.0 | 39.0 |
| Obolon' | 20 | 49.5 | 32.9 | 169.0 | 40.0 |
| Puchezh-Katunki | 80 | 57.1 | 43.6 | 194.0 | 62.7 |
| Gow Lake | 5 | 56.5 | -104.5 | 196.8 | 33.9 |
| Manicouagan | 100 | 51.4 | -68.7 | 215.6 | 22.0 |
| Lake Saint Martin | 40 | 51.8 | -98.5 | 227.8 | 19.2 |
| Paasselkä | 10 | 62.2 | 29.4 | 231.0 | 38.3 |
| Araguainha | 40 | -16.8 | -53.0 | 251.5 | -29.5 |
| Ternovka | 17.5 | 48.0 | 33.1 | 280.0 | 12.7 |
| West Clearwater Lake | 36 | 56.2 | -74.5 | 286.2 | 14.2 |
| Siljan | 52 | 61.0 | 14.9 | 380.9 | -17.2 |
| Nicholson Lake | 12.5 | 62.7 | -102.7 | 387.0 | -20.5 |
| Pilot | 6 | 60.3 | -111.0 | 445.0 | 19.5 |
| Ilyinets | 8.5 | 49.1 | 29.2 | 445.0 | -28.3 |
| Slate Islands | 30 | 48.7 | -87.0 | 450.0 | 2.1 |
| La Moinerie | 8 | 57.4 | -66.6 | 453.0 | 1.3 |
| Kärdla | 4 | 59.0 | 22.7 | 455.0 | -29.5 |
| Tvären | 2 | 58.8 | 17.4 | 458.0 | -32.3 |
| Hummeln | 1.2 | 57.4 | 16.3 | 465.0 | -39.2 |
| East Clearwater Lake | 26 | 56.1 | -74.1 | 465.0 | 3.4 |
| Decorah | 5.5 | 43.3 | -91.8 | 465.5 | 2.9 |
| Granby | 3 | 58.4 | 14.9 | 466.0 | -38.7 |
| Lawn Hill | 18 | -18.7 | 138.7 | 476.0 | 14.7 |
| Carswell | 39 | 58.5 | -109.5 | 481.5 | 17.8 |
| Glikson | 19 | -24.0 | 121.6 | 508.0 | -8 |
| Neugrund | 20 | 59.3 | 23.5 | 535.0 | -43.6 |